\documentstyle[twoside,fleqn,espcrc2,epsf]{article}

\def\case#1#2{{\textstyle \frac{#1}{#2}}}
\def\text#1{{\rm #1}}
\def\dfrac#1#2{{\displaystyle \frac{#1}{#2}}}

\def\vek#1{\mbox{\protect\boldmath $#1$}}
\def\MSbar{{\overline{\rm MS}}}

\newcommand{\AmS}{{\protect\the\textfont2
  A\kern-.1667em\lower.5ex\hbox{M}\kern-.125emS}}

\def\BorderBox#1#2{%
\vbox{\hrule height #1\hbox{\vrule width #1%
	#2 \vrule width #1} \hrule height #1}}
\title{The Charm Quark's Mass%
\hfill\normalsize FERMILAB-CONF-97/326~T}

\author{Andreas S. Kronfeld\address{Theoretical Physics Group, 
        Fermi National Accelerator Laboratory, 
        Batavia, Illinois, U.S.A.}%
        \thanks{Fermilab is operated by Universities Research 
        Association Inc., under contract with the U.S. Dept.\ of 
        Energy.}
        \hfill {\tt hep-lat/9710007}}

\begin{document}

\begin{abstract}
The charm quark's mass is determined from Monte 
Carlo calculations of the $\bar{c}c$ spectrum.
The main sources of uncertainty are perturbation theory (for 
conversion to~$\MSbar$), the continuum-limit extrapolation, Monte Carlo 
statistics, and the effects of quenching.
The (preliminary) result for the $\MSbar$~mass is
$\bar{m}_{\rm ch}(m_{\rm ch})=1.33\pm 0.08$~GeV. 
\end{abstract}

\maketitle

\section{INTRODUCTION}

Nonperturbative lattice calculations of the ha\-dron spectrum provide a 
connection between experimentally measured masses and the couplings of 
the (lattice) QCD Lagrangian.
By convention, however, the $\MSbar$ couplings~$\bar\alpha(M_Z)$ 
and~$\bar{m}(\mu)$, used in phenomenology, are usually quoted.
The two sets of definitions can be related to each other in 
perturbation theory.
For example,
\begin{equation}\label{eq:mm}
\bar{m}(\mu) = M(a)\left[1 + 
	\dfrac{\alpha(q^*)}{4\pi}\left(C_0+\gamma_0\ln\mu^2a^2\right)\right],
\end{equation}
where $\gamma_0=4$.
The lattice mass~$M$ and, by implication, $C_0=C_0(M)$ are specified 
below.
Eq.~(\ref{eq:mm}) omits higher orders in the gauge coupling~$\alpha$ and
power-law artifacts.

This paper determines the charm quark's mass, $\bar{m}_{\rm ch}$, from 
quenched calculations of the $\bar{c}c$ spectrum.
To anticipate the main sources of uncertainty, let us recall recent 
determinations the average of the up and down quarks' 
masses~\cite{Gou97,Gup97}.
There the three largest uncertainties~\cite{Gou97} stem from, in 
descending order, the quenched approximation, the extrapolation to the 
continuum limit (even with the clover action), and perturbation 
theory.
Each of these takes on a different guise for charm, however.

The error in a coupling from quenching can be partly explained by 
noting that couplings run differently in the quenched 
approximation~\cite{Kha92,Mac94}.
One can account for this effect by running the couplings down to 
typical mesonic momenta with $n_f=0$ and then back up to a high scale 
with $n_f\neq 0$.
But~$\bar{m}(\mu)$ does not run for~$\mu<m$,
so quenching should not affect $\bar{m}(m)$ much~\cite{Dav94}.

One might expect lattice spacing errors to be worse for charmonium 
than for light mesons, since $0.4<m_{0,\rm ch}a<1$ on our lattices.
Our spectrum calculations are of mass splittings and the so-called 
kinetic mass of the meson, for which the cutoff effects are powers 
of~$|\vek{p}a|$, not $m_0a$~\cite{KKM97}.
Indeed, we exploit two methods for determining $\bar{m}_{\rm ch}$, 
with opposite cutoff dependence.
The two continuum limits agree, so cutoff effects are under 
{\em better\/} control.

That leaves perturbation theory as the source of the largest 
uncertainty.
To make the most of the one-loop approximation, the only order 
available, we use results for $m_0a\neq0$~\cite{Mer94,Mer97}.
Furthermore, we try to reduce the effect of truncating at one loop by 
choosing~$\alpha(q^*)$ in Eq.~(\ref{eq:mm}) to absorb logarithms 
from higher orders~\cite{BLM83,Lep93}.

\section{CUTOFF EFFECTS}
\label{sect:cutoff}
In a heavy-quark system, such as charmonium, typical three-momenta are 
only a few hundred~MeV, suggesting that worrisome lattice artifacts 
are of order~$(m_0a)^s$.
On the other hand, it is well-known that actions for Wilson fermions 
approach the static limit as $m_0a\to\infty$, showing that
higher-dimension operators are suppressed by a factor of
order~$1/(m_0a)^r$.
The lattice Hamiltonian (defined by the transfer matrix) clarifies the 
middle ground, $m_0a\approx 1$.
One finds~\cite{KKM97}
\begin{equation}\label{eq:artifact}
\hat{H}_{\rm lat}= \hat{H}_{\rm cont} + \delta\hat{H}.
\end{equation}
Contributions to the artifact~$\delta\hat{H}$ take the form
\begin{equation}\label{eq:estimate}
\langle a\delta\hat{H}^{[l]}_n\rangle\sim
g^{2l}b^{[l]}_n(m_0a)|\vek{p}a|^{s_n+1},
\end{equation}
where $\vek{p}$ is a few hundred~MeV, and $s_n>0$.
The function $b^{[l]}_n$ is bounded~\cite{KKM97}.
It is safe to replace it by a number of order unity, and thus the 
effect is about the same size for splittings of charmonium as for 
masses of light-quark hadrons.

Eq.~(\ref{eq:estimate}) applies only if the hopping parameter~$\kappa$ 
is adjusted until the meson's kinetic mass
\begin{equation}
M_2:=(\partial^2E/\partial p_i^2)^{-1}_{p_i=0}
\end{equation}
equals the meson's physical mass.
When $Ma\neq 0$, the rest mass $M_1:=E({\bf 0})$ is smaller.
Nevertheless, the {\em splittings\/} of meson rest masses are accurate 
up to Eq.~(\ref{eq:estimate}).
In particular, the spin-averaged binding energy
\begin{equation}\label{eq:B1}
B_1a := (M_{1\bar{Q}Q}a)^{\text{MC}} - 2(M_{1Q}a)^{\text{PT}},
\end{equation}
where $M_{1\bar{Q}Q}$ is the spin average of mesons' rest masses,
has relative errors of order $\min(\vek{p}^2a^2,v^2)$ \cite{Kro97}.
(When the quark's rest mass~$M_{1Q}$ is computed to finite order in 
perturbation theory, $B_1a$ suffers perturbative errors as well.)

To determine $\bar{m}_{\rm ch}$ we rely, therefore, on the following
Monte Carlo calculations:
We define the lattice spacing~$a$ in physical units from
$\Delta M=M_{h_c}-\case{3}{4}(M_{\eta_c}+3M_{J/\psi})$~\cite{Kha92}.
We then obtain the quark mass either from the spin-averaged binding 
energy~$B_1$ of the 1S states, or from their spin-averaged kinetic 
mass~$M_{2\bar{Q}Q}$.

\section{PERTURBATION THEORY}
\subsection{When $m_0a\neq 0$}
If the Monte Carlo has~$m_0a\neq 0$ it is necessary to 
take~$m_0a\neq 0$ when deriving Eq.~(\ref{eq:mm}).
Although $C_0$ remains bounded~\cite{KKM97,Mer94,Mer97}, its value can 
change significantly for nonzero~$m_0a$.

Eq.~(\ref{eq:mm}) is obtained by computing the quark's pole mass in 
lattice and in $\MSbar$ perturbation theory.
Because the lattice breaks Euclidean invariance, several ``masses'' 
($M_1$, $M_2$, etc) describe the pole.
One would like to pick a pole mass without dire lattice 
artifacts.
We use two methods.
In the first, we take the binding energy and set
\begin{equation}\label{eq:mp1}
m_{\text{pole}} =
\case{1}{2}(M_{\bar{Q}Q}^{\text{expt}}-B_1)
\end{equation}
with $B_1a$ from Eq.~(\ref{eq:B1}) and $a$ from $\Delta M$.
In the second method, we use the quark's kinetic mass, but reduce 
\pagebreak[3]
uncertainty in tuning~$\kappa$ by taking~$a$ from the meson's 
kinetic mass:
\begin{equation}\label{eq:mp2}
m_{\text{pole}} = (M_{2Q}a)^{\text{PT}}
\frac{M_{\bar{Q}Q}^{\text{expt}}}{(M_{2\bar{Q}Q}a)^{\text{MC}}}.
\end{equation}
When $B_1a$ and $M_{2Q}a$ are expanded in perturbation theory, 
Eqs.~(\ref{eq:mp1}) and~(\ref{eq:mp2}) can be matched to the expansion 
of~$m_{\rm pole}$ in~$\MSbar$.
The manipulations at one loop define $M$ and $C_0$ in Eq.~(\ref{eq:mm}).

One needs, therefore, the loop corrections to the quark's rest and 
kinetic masses.
  From formulas~\cite{Mer94,Mer97} for $M_1$ and $M_2$, to all orders 
in~$g_0^2$ and in~$m_0a$, one can expand
\begin{equation}
M_1=\sum_{l=0}g_0^{2l}M_1^{[l]}.
\end{equation}
One finds $M_1^{[0]}=\log(1+M_0)$, where 
$M_0=1/2\kappa-1/2\kappa_{\rm crit}$.
Refs.~\cite{Mer94,Mer97} show results for~$M_1^{[1]}$.

The kinetic mass has further loop corrections,
so let $Z_{M_2}(M_1)=M_2/m_2(M_1)$.
The function $m_2(M)$ is chosen so that $Z_{M_2}^{[0]}=1$, but it is 
evaluated at the all-orders~$M_1$.
With this definition $Z_{M_2}(0)=1$, to all orders in~$g_0^2$.
Also, $Z_{M_2}^{[1]}$ is tadpole-free.
It is small ($0\ge Z_{M_2}^{[1]}>-0.1$) and hardly depends on the clover 
coupling~$c_{\rm SW}$~\cite{Mer97}.

\subsection{Choosing $\alpha(q^*)$}
\label{sect:q}
With only the one-loop approximation at hand, the right-hand side of
Eq.~(\ref{eq:mm}) is sensitive to the choice of scheme for~$\alpha$ and
its scale~$q^*$.
Since Eq.~(\ref{eq:mm}) is the combination of lattice and $\MSbar$
perturbation theory, the original series must be expressed in a common
schem and the scales must be run to a common one.
Here we use the scales suggested in Refs.~\cite{BLM83,Lep93}, primarily
for~$\alpha_V$, but also for~$\bar\alpha$.

For dimensional regulators Ref.~\cite{BLM83} prescribes
\begin{equation}
\ln q^2_{\rm BLM}/\mu^2 = I^*/I,
\end{equation}
where $I^*$ is derived from the Feynman diagram for $I$ by replacing 
gluon propagators by
\begin{equation}
q^{-2}\mapsto 
q^{-2}\left[\ln(q^2/\mu^{2}) - b_1^f/\beta_0^f\right].
\end{equation}
The constant depends on the scheme: 
for $\bar{\alpha}$, $b_1^f/\beta_0^f=5/3$;
for $\alpha_V$, $b_1^f/\beta_0^f=0$.

Similarly, for the lattice Ref.~\cite{Lep93} prescribes 
\begin{equation}
\ln q^2_{\rm LM}a^2 = I^*/I,
\end{equation}
where $I^*$ now comes from the replacement
\begin{equation}
\hat{q}^{-2}\mapsto 
\hat{q}^{-2}\ln(\hat{q}^2a^2).
\end{equation}
With no constant, this prescription is for the coupling defined in 
Ref.~\cite{Lep93}, which coincides with~$\alpha_V$ through
next-to-leading order.
\begin{figure} 
\BorderBox{0pt}{
	\epsfxsize=0.46875\textwidth
	\epsfbox{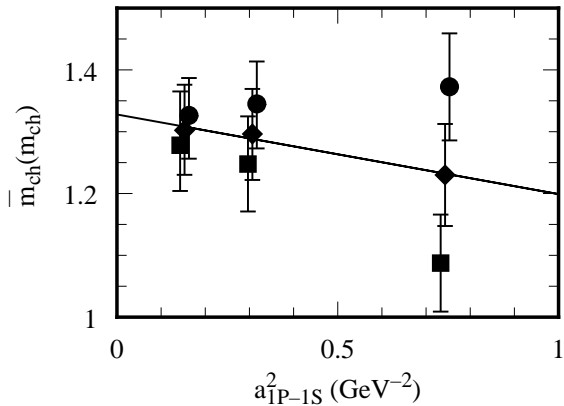}
}
\vspace*{-5mm}
	\caption{The charm quark's $\MSbar$ mass vs.~lattice spacing
	squared.
	Circles (squares) denote Method 1 (2).
	The curve is a linear fit of the average (diamonds).
	Offsets in $a^2$ are for clarity.}
	\label{fig:mCharm}
\end{figure}

When combining the series to form Eq.~(\ref{eq:mm}), one can 
combine~$q^2_{\rm BLM}$ and~$q^2_{\rm LM}$ in the usual way,
\begin{equation}
\ln(q^{*2}a^{2})=
(I^*_{\rm lat}-I^*_{\rm cont})/(I_{\rm lat}-I_{\rm cont})
\end{equation}
provided the constants used to define the $I^*$s are compatible.
(Otherwise the final $q^*$ has problems as $ma,\,m/\mu\to 0$.)
Most straightforward, we find, is to use $\alpha_V$ and
to extract $\bar{m}_{\rm ch}(m_{\rm ch})$ directly from 
Eq.~(\ref{eq:mm}).
The resulting $q^*$s are a few~GeV but somewhat $a$~dependent.

\section{RESULTS}
We have computed the charmonium spectrum for
$(\beta,~c_{\rm SW})=(5.5,~1.69)$, (5.7,~1.57), (5.9,~1.50), 
and~(6.1,~1.40)~\cite{Kha98}.
Our (preliminary) results for $\bar{m}_{\rm ch}(m_{\rm ch})$ with 
tadpole-improved perturbation theory are plotted against~$a^2$ in 
Fig.~\ref{fig:mCharm}.
The error bar is dominated by the unknown two-loop correction to
Eq.~(\ref{eq:mm}), estimated to be twice the
square of the one-loop term.
When the analysis is repeated without tadpole improvement, but still
choosing $\alpha(q^*)$ as in Sect.~\ref{sect:q}, the data change
negligibly.
The subdominant uncertainty is from the Monte Carlo statistics
of~$M_{2\bar{Q}Q}$.

Extrapolating the average of the two methods linearly in~$a^2$ yields
\begin{equation}
\bar{m}_{\rm ch}(m_{\rm ch}) = 1.33 \pm 0.08~{\rm GeV}.
\end{equation}
The error bar now incorporates uncertainty in the extrapolation,
e.g., extrapolating linearly in~$a$.
Note that the quoted result neither explicitly corrects for, nor 
assigns an error to, quenching, because~$\bar{m}(\mu)$ does not run
when $\mu<m$~\cite{Dav94}.

A 6\% uncertainty for the charm quark's mass is twice the~3\% quoted
for the top quark's mass from collider experiments.
Alas, without two-loop (or nonperturbative) matching for $m_0a\neq0$,
top standards will be impossible to achieve.

\vspace*{3mm}
This work was carried out with A. El-Khadra, P. Mackenzie,
B. Mertens, and J. Simone.

\end{document}